\documentclass[fleqn,twoside,twocolumn,nofootinbib]{revtex4} 
\usepackage{ujp} 
\begin{document}

\title[NATURE OF HYDROGEN BOND IN WATER]
{NATURE OF HYDROGEN BOND IN WATER}%

\author{P.V. Makhlaichuk, M.P. Malomuzh, I.V. Zhyganiuk  }
\affiliation{I.I. Mechnikov Odessa National University, Dept. of Theoretical Physics}
\address{2, Dvoryanska Str., Odessa 65026, Ukraine}
\email{mahlaichuk@gmail.com}

\udk{???}  \razd{\secix}

\setcounter{page}{113}%

\maketitle

\onecolumngrid \vspace{-3mm}
\begin{flushright}
\textit{In memory of Professor Galina Puchkovskaya}
\end{flushright}
\vspace{3mm} \twocolumngrid

\begin{abstract}
The work is devoted to the investigation of the physical nature of
H-bonds. The H-bond potential $\Phi _{H} (r,\Omega )$  is studied as
an irreducible part of the interaction energy of water molecules. It
is defined as a difference between the generalized Stillinger--David
potential and the sum of dispersive and multipole interaction
potentials. The relative contribution of $\Phi _{H} (r,\Omega )$  to
the intermolecular potential does not exceed $(10\div15)\%$.
\end{abstract}

\section{Introduction}
The simplest structure of intermolecular interaction potential is
characteristic of the atomic liquids such as argon. Their
intermolecular interaction potential $\Phi (r)$ is the sum of the
attractive part $\Phi _{\rm dis} (r)$ caused by dispersive forces
and $\Phi _{\rm rep} (r)$ that describes the repulsion:
\begin{equation} \label{eq1}
\Phi (r)=\Phi _{\rm rep} (r)+\Phi _{\rm dis} (r).
\end{equation}
In particular, the well-known Lennard-Jones potential has such a
structure. For molecules having no spherical symmetry, the
intermolecular potential becomes angular dependent \cite{Krokston}:
\begin{equation} \label{eq2}
\Phi (r)\to \Phi (r,\Omega )=\Phi _{\rm rep} (r,\Omega )+\Phi _{\rm
dis} (r,\Omega ),
\end{equation}
where $\Omega$   denotes the set of angles which describe the
relative orientation of molecules. Such form of the intermolecular
potential is characteristic of molecules N$_2$. For molecules
without center of inversion, it is necessary to consider the
dipole-dipole interaction and multipole interactions of higher
order. This circumstance leads to the additional term
$\Phi_{M}(r,\Omega)$:
\begin{equation} \label{eq3}
\Phi (r,\Omega )=\Phi _{\rm rep} (r,\Omega )+\Phi _{\rm dis}
(r,\Omega )+\Phi _{M} (r,\Omega ).
\end{equation}
The analogous structure of interaction potential is also inherent to
molecules in water and water-alcohol solutions if their electronic
shells do not overlap. At small distances between molecules, the
overlapping effect of the electronic shells becomes essential. The
corresponding interaction is usually called the hydrogen bond
(H-bond). The intermolecular potential is represented in the form
\cite{Ei-Kau}
\begin{equation} \label{eq4}
\Phi (r,\Omega )=\Phi _{\rm dis} (r,\Omega )+\Phi _{H} (r,\Omega ),
\end{equation}
where $\Phi _{H}(r,\Omega )$ includes also the repulsion and multipole interactions.

From the qualitative point of view, such a change of priorities is not justified,
since the analytic continuation of multipole contributions to the overlapping region
does not lead to the effects that can violate the continuity condition of the potential.
That is why we should redefine the H-bond potential. In accordance with this,
the H-bond potential will be defined as
\begin{equation} \label{eq5}
\Phi (r,\Omega )=\Phi _{\rm rep} (r,\Omega )+\Phi _{\rm dis}
(r,\Omega )+\Phi _{M} (r,\Omega )+\Phi _{H} (r,\Omega ),
\end{equation}
where $\Phi_{dis}(r,\Omega)$ and $\Phi_{M}(r,\Omega)$ are the analytic
continuations in the overlapping region of electronic shells.

One  of the most characteristic  manifestations of H-bonds in liquid
water and its  vapor is the formation of dimers and multimers of
higher order. In other words, the properties of a dimer give us the direct
information about H-bonds. This fact points a way how to study the
intermolecular interaction in water and, rigorously speaking, the
formation of H-bonds. Let us note the main steps of this approach:
1)~the interaction energy of two water molecules is described with
the help of the most suitable phenomenological model potential that
describes the ground state of a dimer; 2)~the energy obtained in such
a way is compared to that for a water dimer calculated with the help
of the asymptotic multipole expansion; 3)~to determine the H-bond
potential, we construct the difference between the model potential
and the sum of dispersive and multipole contributions. We expect
that this difference will be non-zero only in a small region near
the equilibrium distance between the water molecules in
a dimer.

We suppose that the generalized Stillinger--David potential (GSD)
\cite{Zhyg} is the most suitable intermolecular potential for the
description of the interaction between water molecules. This is a
soft potential, whose parameters can change under the influence of
nearby molecules. This important circumstance cannot be taken into
account for almost all phenomenological potentials
\cite{Jorg-Chan,Anton-Davyd,Bern-Fowl,Rahm-Stil,Mal-Polt}. In
contrast to the original work \cite{SD}, the role of the screening
functions (or overlapping effects) in GSD is taken into account
more adequately. Moreover, the asymptotic behavior of the original
Stillinger--David  potential (SD) is corrected for large distances
between molecules.

\begin{figure}
\includegraphics[width=8cm]{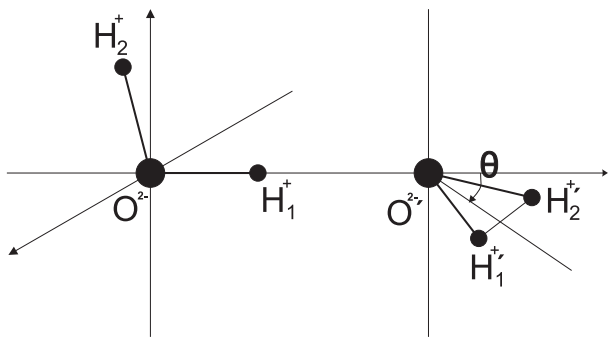}
\vskip-3mm\caption{Equilibrium configuration of a water
dimer}\vskip3mm
\end{figure}

\begin{figure}
\includegraphics[width=8cm]{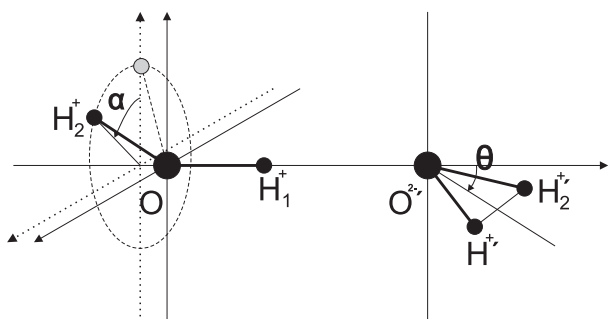}
\vskip-3mm\caption{Definition of angle $\alpha$}
\end{figure}

The  multipole moments (up to octupole) can be determined by the
quantum-chemistry methods with sufficient accuracy [10, 11]. Due to
this, we are able to construct the asymptotic estimation of the
interaction potential for distances larger than the screening
radius. At the same time, the analytic continuation of the multipole
potential to the overlapping region does not lead to any serious
errors. This allows us to construct the above-mentioned difference
between the suitable model potential and multipole
contribution.

The  main goal of this work is the realization of the program
formulated above for the construction of the H-bond potential.

\section{Ground State of Water Dimer}
In this section, we will represent the results of our study of
dimers with the help of the GSD potential [3]. In order to
facilitate the calculations, we will use the hard model of water
molecules (i.e., the position of hydrogens and oxygen remain fixed,
as well as their configuration). According to [3],  such a
requirement leads to the error of $(1.5\div3)\%$. The ground state
of a dimer is identified to the minimum of the interparticle
potential for two molecules presented in Fig.~1.

The equilibrium distance $r_{\rm OO}^{(0)}$ between oxygens and
the angle $\theta_{0}$ between the dipole moments of a water
molecule are determined from the absolute minimum of the GSD
potential: $\min \Phi_{\rm GSD}(\tilde{r},\theta,\alpha=0)$. It is
considered as a function of the dimensionless distance
$\tilde{r}=r_{\rm OO}/r_{\rm OH}$, where $r_{\rm OH}=0.97$~{\AA}
is the distance between the oxygen and a hydrogen in a water
molecule, the angle $\theta$ describes some bend of the H-bond,
and $\alpha$ is related to the rotation around the H-bond (see
Fig.~2).

\begin{figure}
\includegraphics[width=\column]{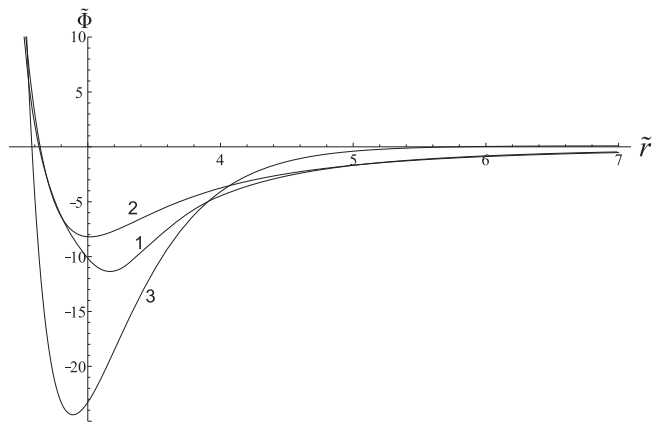}
\vskip-3mm\caption{Comparative behavior of GSD potential ({\it 1}),
multipole potential ({\it 2}), and SD potential ({\it 3})}
\end{figure}

The comparative behavior of $\Phi_{\rm GSD}$, multipole potential
$\Phi_{\rm MP}$ that will be studied below, and $\Phi_{\rm SD}$ from
the original work \cite{SD} is presented in Fig. 3. The
equilibrium values of the distance $r_{\rm OO}^{(0)}$ between the
oxygens in a dimer, the angle $\theta_{0},$ and the ground-state energy
$\tilde{E}_{d}$ ($\tilde{E}_{d}=E_{d}/k_{\rm B} T_m$, $T_m$ is the
melting (crystallization) temperature for liquid water, $k_{\rm B}$
is the Boltzmann constant) for all potentials investigated are
presented in Table. In addition, the dipole moment $D_d$ of a dimer is
included to Table as well. It is determined according to the formula
\begin{equation} \label{eq7}
D_{d} =2d_{w} \cos \frac{1}{2} \left(\theta _{0} +\delta
/2\right)\approx \sqrt{2} d_{w} \approx 2.6{\rm  D}.
\end{equation}

\begin{table}[b]
\noindent\caption{Equilibrium parameters of water
dimer}\vskip3mm\tabcolsep7.2pt

\noindent{\footnotesize
\begin{tabular}{c c c c c}
 \hline%
 \multicolumn{1}{c}{\rule{0pt}{9pt}Dimer parameters}%
 & \multicolumn{1}{|c}{$r_{\rm OO}^{(0)}$, {\AA} }
 & \multicolumn{1}{|c}{$\theta_{0}$, deg}
 & \multicolumn{1}{|c}{$E_{d}$}
 & \multicolumn{1}{|c}{$d_{d}$, D}\\%
\hline%
\rule{0pt}{9pt}GSD&3.07&31.1&--11.36&2.76\\ 
GSD($\lambda=0.98$)&3.00&18.2&--13.8&3.02\\%
$\Phi_{\rm MP}(SD)$&3.00&26.8&--8.73&3.23\\%
$\Phi_{\rm MP}(SP)$&3.01&27.9&--8.19&3.08\\%
\cite{Umey-Moro}&2.98&60&--10.69&\\%
\cite{Schutz}&2.925&47.5&--9.11&\\%
\cite{Matsuoka}&&&--10.32&\\%
\cite{Odutola}&2.976&57 $\pm$ 10&&\\%
\cite{Yu}&&&&2.6\\%
\hline
\end{tabular}
}
\end{table}

Here, $d_w$ denotes  the modulus of the water molecule dipole
moment,  $\delta$ is the equilibrium angle between lines connecting
oxygen and hydrogens. The second line in Table corresponds to
the GSD potential, in which the screening length for the function
$1-L(r)$ (see \cite{SD}) is decreased by $\lambda=0.98$ times. The
dependence of the interaction energy on the rotation angle $\alpha$
around the H-bond is especially important (see Fig.~4).

As is seen, the inequality $\Delta \tilde{\Phi}(\alpha)<1$, $\Delta
\tilde{\Phi}(\alpha)=\frac{\Phi(\tilde{r}_{\rm
OO},\theta_0,\alpha)-\Phi(\tilde{r}_{\rm OO},\theta_0,0)}{k_{\rm B}
T_m}$, indicates the possibility of intramolecular rotations of
water molecules in a dimer. This circumstance should manifest
itself in the entropy and heat capacity behavior \cite{PLA2}.

\section{Multipole approximation for the interparticle potential}

The  usage of the multipole interaction potential is justified by
the following reasons: 1) multipole interaction potential of two
water molecules is more convenient in specific calculations; 2)
determination of multipole moments within the methods of quantum
chemistry is much simpler than the construction of approximating
functions; 3) comparison of the attractive part of model potentials
and the electrostatic multipole  potential for the dimer
configuration allows us to control the degree of applicability of model
potentials.

The  last requirement, as it will be shown below, leads to the
conclusion that the GSD potential is the most suitable for the
description of water dimers.

The multipole potential of the intermolecular interaction is modeled as
\begin{equation}\label{eq9}
\Phi_{\rm MP}(r)=\Phi_{0}\exp(-k(r-\sigma))+\Phi_{M}(r),
\end{equation}
where the repulsion part will have the same behavior as that for the GSD
\cite{Zhyg}. Here, $\sigma$ is the doubled hard-core radius of a
water molecule, which is identified as the H-bond length
\cite{Ei-Kau}, and $\Phi_M(r)$  is a part of the multipole expansion
for the interaction energy between two water molecules.

\begin{figure}
\includegraphics[width=\column]{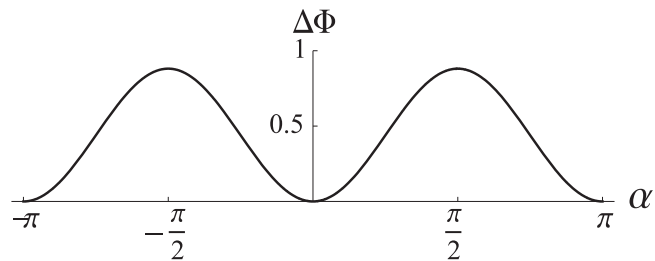}
\vskip-3mm \caption{$\alpha$-dependence of the interaction
energy for a dimer at $r_{\rm OO}=3$ {\AA} }
\end{figure}

We use the multipole expansion $\Phi_M(r)$ up to the quadrupole-quadrupole and dipole-octupole terms:
\begin{equation} \label{eq9}
\Phi _{M} (r)=\Phi _{DD} +\Phi _{DQ} +\Phi _{QQ} +\Phi _{DO}
+\ldots\,.
\end{equation}

For the dimer configuration in Fig.~1, the corresponding terms are
\[
\Phi_{DD} =\frac{d^{2} }{r_{\rm OO}^{3} } (\cos (\delta /2+\theta
)-3\cos \delta /2\cos \theta ),
\]
\[
\Phi_{DQ}=-\frac{1}{2 r_{\rm
OO}^{4}}(6\{d_{\alpha}^{(1)}Q_{2\alpha}^{(2)}+d_{\alpha}^{(2)}Q_{2\alpha}^{(1)}\}
- \]
\[ - 15\{d_{2}^{(1)}Q_{22}^{(2)}+d_{2}^{(2)}Q_{22}^{(1)}\}),
\]
\[
\Phi_{QQ}=\frac{3}{4 r_{\rm OO}^{5}}(35 Q_{22}^{(1)} Q_{22}^{(2)} -
20 Q_{2\delta}^{(1)} Q_{2\delta}^{(2)} + Q_{\delta \gamma}^{(1)}
Q_{\delta \gamma}^{(2)}),
\]
\[
\Phi_{DO}=-\frac{3}{2 r_{\rm OO}^5}(15 \{d_{2}^{(1)} O_{222}^{(2)} +
d_{2}^{(2)} O_{222}^{(1)}\} - \]
\[ - 5 \{d_{\alpha}^{(1)} O_{22
\alpha}^{(2)} + d_{\alpha}^{(2)} O_{22 \alpha}^{(1)}\}).
\]

The  analysis of the repulsive part of the multipole potential shows
that the parameters  $\Phi_0$, $k$, and $\sigma$ take values
$\Phi_0=8.5$, $k=5$, and $\sigma=2.8$. The relative values of different
contributions to $\Phi_M$  are shown in Fig.~5. Lines ({\it 3})
and ({\it 4}) testify that the quadrupole-quadrupole and
dipole-octupole interactions have the same order of magnitude.

As is seen from Fig.~3, the multipole part is able to correctly
describe the interaction energy at distances up to $4$~{\AA}.
More exactly, on the interval $4$~{\AA}~$<r_{\rm OO}<\inf,$ the
values of multipole potential and GSD coincide with high
accuracy. At smaller distances between oxygens, the overlap of
electronic shells begins. The multipole approach becomes
inapplicable. At the same time, the phenomenological model potentials
are supposed to be used also in this region. In particular, the
applicability of the GSD potential within the overlapping region is
justified by a suitable selection of the screening functions.
Unfortunately, the most of other phenomenological potentials have no
necessary compliance. The generalized Stillinger--David potential,
used in our consideration of dimer properties, is quite satisfactory,
and it has the ability for further modifications.

\begin{figure}
\includegraphics[width=\column]{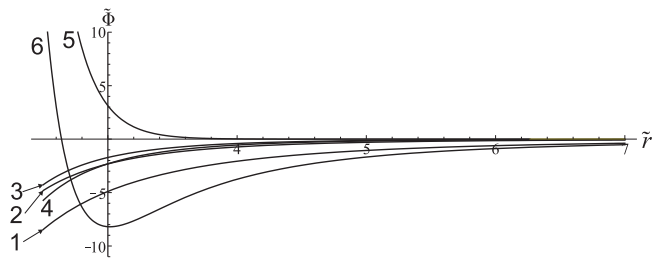}
\vskip-3mm\caption{Role of the terms $\Phi _{DD} ,\, \Phi _{DQ} ,\,
\Phi _{QQ} $, and $\, \Phi _{DO} $ in the multipole interaction
potential (7)  of two water molecules in the dimer configuration:
{\it  1} -- $\Phi _{DD} $, {\it 2} -- $\Phi _{DQ} $, {\it 3} --
$\Phi _{QQ} $, {\it 4} -- $\, \Phi _{DO} $, {\it 5} -- $\Phi_{\rm
rep}$, {\it 6} -- $\Phi_{\rm MP}$ }\vskip3mm
\end{figure}

\section{Hydrogen Bond Potential}

According to the definition of H-bond given in Introduction, we will
consider the difference $\Phi_{H}(r)$ between the generalized
Stillinger--David potential $\Phi_{\rm GSD}$ and the sum of the
multipole potential $\Phi_{M}$ and the dispersive energy
\cite{Mal-Polt}:
\begin{equation}
\Phi _{H} (r,\Omega )=\Phi _{\rm GSD} (r,\Omega )-\Phi _{r}
(r,\Omega ) -\Phi _{\rm dis} (r,\Omega )-\Phi _{\rm MP} (r,\Omega ).
\end{equation}
The behavior of the potentials $\Phi_{\rm GSD}(r,\Omega)$,
$\Phi_{\rm dis}(r,\Omega),$ and $\Phi_{\rm MP}(r,\Omega)$ for the
configuration of molecules characteristic of a dimer is presented
in Fig.~6.

\begin{figure}
\includegraphics[width=\column]{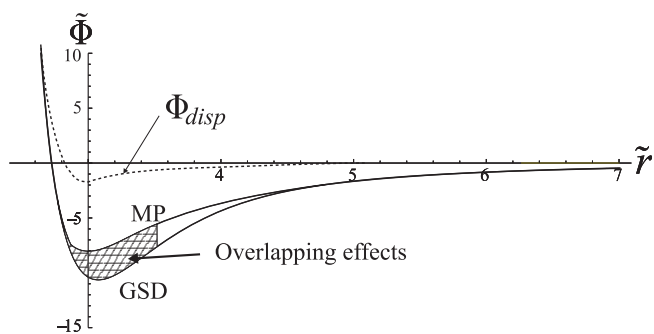}
\vskip-3mm\caption{Partial contributions to the intermolecular potential.}
\end{figure}

The H-bond potential for the same configuration is presented in
Fig.~7.

\begin{figure}
\includegraphics[width=\column]{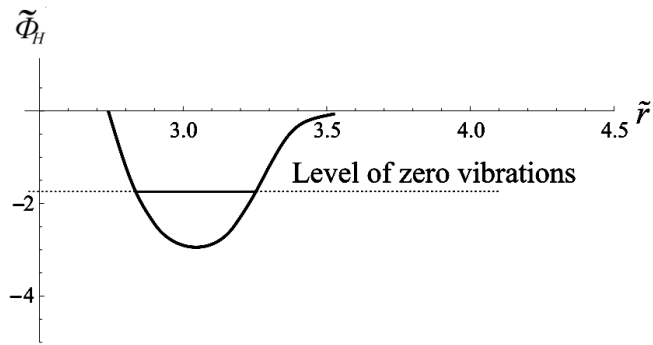}
\vskip-3mm\caption{H-bond potential.}
\end{figure}

Hence, $\Phi_H(r,\Omega)$ is a short-range potential that
appears due to the overlapping of the electronic shells, and it has the
quantum mechanical nature. It should be interpreted as an H-bond
potential in water. It takes the same order of magnitude as the
dispersive term and is much smaller than the multipole
interaction ($\Phi_{\rm dis}~(1\div2)k_{\rm B} T_m$, $\Phi_{\rm
MP}~(7\div8)k_{\rm B} T_m$, $\Phi_{H}~(2\div3)k_{\rm B} T_m$). That
is why the contribution of H-bonds to the thermodynamic potential
can be taken into account with the help of perturbation theory. This
circumstance is confirmed qualitatively by the similarity of
thermodynamic properties of water and argon on their coexisting
curves \cite{CPL2}.

\section{Discussion}

The relatively small depth of the potential well of a hydrogen bond
leads to the following conclusions: 1) contributions of
$\Phi_H(r,\Omega)$ to the thermodynamic potentials and the kinetic
coefficients can be calculated with the help of perturbation theory;
2) temperature behavior of the thermodynamic characteristics such as
the fraction volume or the heat of evaporation is argon-like with
satisfactory accuracy. The last conclusion is confirmed by the
results from [18--20]. The H-bond potential in Fig.~7 corresponds to
the equilibrium orientation of water molecules. There are no
restrictions  to construct $\Phi_H(r,\Omega)$  for all other
orientations: all angular dependences of $\Phi_{\rm GSD}(r,\Omega)$
and  $\Phi_{\rm MP}(r,\Omega)$  are well known. However, we have to
mention that the thermodynamic properties of water are defined by
the averaged intermolecular potential due to the rotation of water
molecules [20]. It is shown [20] that the potential well depth
reduces due to the averaging over angular variables. This leads to
the correction of the argon-like dependences of thermodynamic
quantities by at most $5\%$ [18, 20]. Nevertheless, the
contributions of H-bonds exhibit themselves in the heat capacity of
water \cite{PLA2}, dipole relaxation, and spectral properties
\cite{Ei-Kau}. Another important circumstance is the necessity to
consider the influence of the neighbors on the H-bond potential.
This collective effect will be studied in details in further
publications.

\vskip3mm The authors cordially thank Prof. V. Balevicius, Prof. L.
Kimtys, Prof. A. Koll, Prof. A.V. Kraiski, Prof. N.I.~Le\-bovka,
Prof. L.N. Lisetskiy, Prof. N.N.~Mel\-nik, and especially Prof. V.E.
Pogorelov for the fluidful discussion of the results obtained.

\rezume{ПРИРОДА ВОДНЕВОГО ЗВ'ЯЗКУ У ВОДІ}{П.В. Махлайчук, М.П.
Маломуж,  І.В. Жиганюк} {Роботу присвячено дослідженню фізичної
природи водневого зв'язку, який утворюється між молекулами води.
Потенціал водневого зв'язку $\Phi_H(r,\Omega)$  розглядається як
незвідна частина енергії взаємодії молекул води, що визначається
різницею між узагальненим потенціалом Cтілінджера та Девіда і сумою
потенціалів дисперсійної та мультипольної взаємодій. Показано, що
відносна величина внеску  $\Phi_H(r,\Omega)$ не перевищує
(10--15)\%.}

\end{document}